# COVID-VIT: Classification of Covid-19 from CT chest images based on vision transformer models

Xiaohong Gao, Yu Qian, Alice Gao*

Department of Computer Science, Middlesex University, London, UK.

x.gao@mdx.ac.uk.

* A&E Department, Newham University Hospital, Barts Health NHS Trust, London, UK.

alicegao@doctors.org.uk.

**Abstract**

This paper is responding to the MIA-COV19 challenge to classify COVID from non- COVID based on CT lung images. The COVID-19 virus has devastated the world in the last eighteen months by infecting more than 182 million people and causing over 3.9 million deaths. The overarching aim is to predict the diagnosis of the COVID-19 virus from chest radiographs, through the development of explainable vision transformer deep learning techniques, leading to population screening in a more rapid, accurate and transparent way. In this competition, there are 5381 three-dimentaional (3D) datasets in total, including 1552 for training, 374 for evaluation and 3455 for testing. While most of the data volume are in axial view, there are a number of subjects' data are in coronal or sagittal views with 1 or 2 slices are in axial view. Hence, while 3D data based classification is investigated, in this competition, 2D images remains the main focus. Two deep learning methods are studied, which are vision transformer (ViT) based on attention models and DenseNet that is built upon conventional convolutional neural network (CNN). Initial evaluation results based on validation datasets whereby the ground truth is known indicate that ViT performs better than DenseNet with F1 scores being 0.76 and 0.72 respectively. (Codes are available at GitHub at https://github.com/xiaohong1/COVID-ViT).

## 1. Introduction

COVID-19, officially known as SARS-CoV-2 is a strain of coronavirus. The first cases were seen in Wuhan, China, in late December 2019 before spreading globally [1-3] which as and was classified as a pandemic in March 2020 [4]. At present there are more than 182 million people infected with the virus and 3.9 million of deaths [5] with new variants keep appearing.

The clinical picture can range from a mild common cold-like illness, to a severe viral pneumonia leading to acute respiratory distress syndrome (ARDS) that is potentially fatal. The presence of COVID-19 in respiratory specimens was detected by next generation sequencing or real-time reverse transcription polymerase chain reaction (RT-PCR) methods, a laboratory technique combining reverse transcription of Ribonucleic acid (RNA) into Deoxyribonucleic acid (DNA) and amplification of specific DNA targets. While PCR tests offer many advantages, results are not usually available for at least several hours. On the other hand, high resolution Computerised Tomography (CT) are non-invasive, easy to operate and prevalent and hence can assist diagnosis for COVID-19 rapidly.

With regard to imaging features, it appears that bilateral infiltrates with peripheral opacities and patchy consolidation are the most common findings on chest radiographs (CXR) [6,7] and bilateral ground glass opacities is often a key finding on CT [8,9].

As confirmed cases continues to increase considerably all over the world, timely detection of the disease not only can provide supportive care required by patients but also can prevent further spread of the virus. Consequently, effective screening of infected patients appears to be a critical step in this fight against COVID-19 as well as to circumvent the temporary shortage of RT-PCR kits to confirm COVID-19 infection.

The challenge here facing detecting COVID-19 based on chest CT images is that when the disease is at its early onset, the characteristic patterns present less obvious to the human eyes [10]. Hence, machine learning based approaches are applied to investigate COVID-specific biomarkers. In this study vision transformer architectures are investigated.

## 2. Related work

Vision transformer (ViT) have recently demonstrated its potentials in image processing by achieving comparable results while requiring fewer computational resources. Based on self-attention architectures, transformer becomes the leading model in natural language processing (NLP) [11]. For NLP, by employing attention models, i.e. transformers, training speed can be significantly improved hence enhancing the performance of neural machine translation applications. For image processing, vision transformers are emerging and starting to show protentials by applying to computer vision tasks, such as image recognition [12]. Specifically, ViT appears to demonstrate excellent performance when trained on sufficient data, outperforming a comparable state-of-the-art CNN with four times fewer computational resources.

One of the advantages that that Transformers present is computational efficiency and scalability. It has become possible to train models of unprecedented size, with over 100 billion parameters [13].

Figure 1 illustrates the architecture of ViT employed in this study. In this study, the ViT is implemented in pytorch and heavily based on the code at [14].

The training process takes place at a GPU sever that equipped with one Quadro RTX 8000 GPU and 64GB memory under Debian Linux operating system. While in the training, the 2D images are resized to 224×224×3 and 224×224×32 for 3D. For 3D training, each subject's 2D slices (in JPG format) are firstly converted into Analyze (7.5) format, with both header (.hdr) and image (.img) files. The patch size for the application of ViT model is 7×7 for 2D images and 8×8×8 for 3D volumes. It takes about 24 hours for training 80 epochs for 2D images and ~30 hours for 3D volumes.

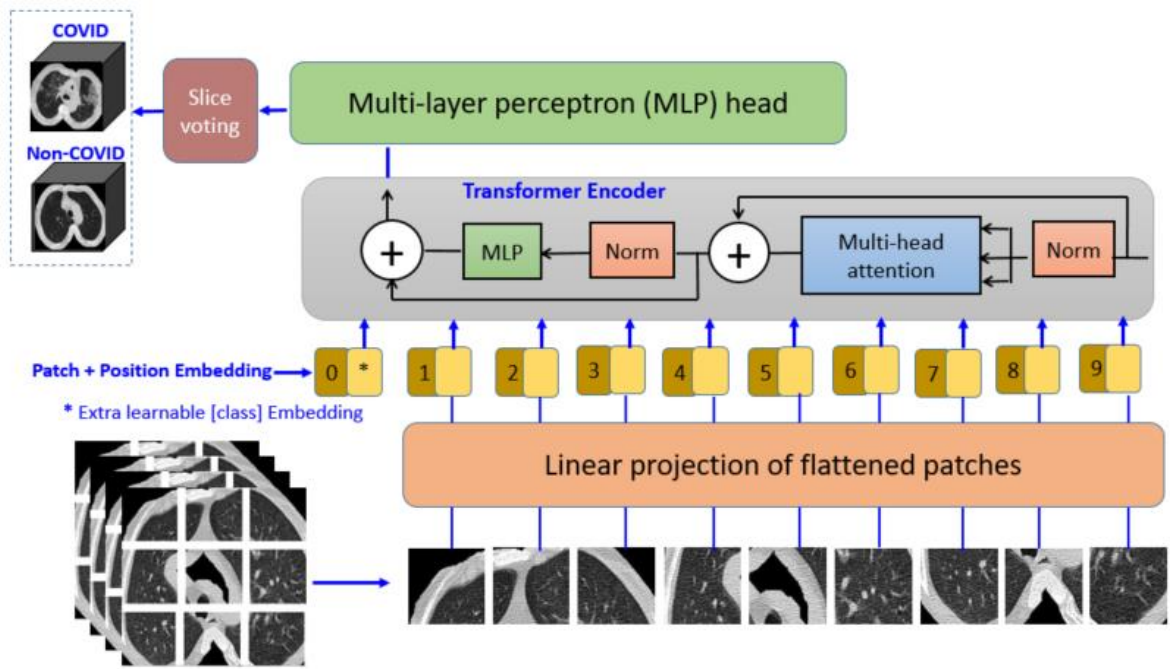

Figure 1. The ViT architecture implemented in this work.

## 3. The MIA-COV19 competition datasets

The CT thorax lung images are collected from MIA-COV19 competition, composing of training, validation and testing datasets. Table 1 lists the detailed information of these data. The resolution of the these images is either 512×512 or 768×768 pixels whereas the depth of each volume ranges from 4 slices to 1026 as detailed in MIA-COV19 competition papers [15-18].

Table 1. The datasets from MIA-COV19 competition applied in this paper. Note 2D slice numbers are the slices that have undertaken pre-processing stage and removed those with little lung contents.

| Label | Train | | Validation | | Testing | | Total |
|---|---|---|---|---|---|---|---|
| | 3D subject | 2D slice | 3D subject | 2D slice | 3D subject | 2D slice | |
| COVID | 687 | 63,808 | 165 | 13,839 | | | |
| Non-COVID | 865 | 66,880 | 209 | 15,793 | | | |
| total | 1,552 | 130,688 | 374 | 29,632 | 3,455 | 444,524 | 5381 |

### 3.1 Image pre-processing

Because the diseased regions of a COVID-19 dataset occupy less than 10% of the whole volume and present in a number of slices, image pre-processing take place first to maximise the large visibility of diseased slices while removing scanner artefact. Figure 2 demonstrates a montage view of all the slices for one subject. It shows that the first 3 slices hardly depict any lung content whereas the boundary information as well as the background in each slice accommodates more than half of the slice in concern in each 2D image. In addition, the heart (arrow) and liver (arrow head) also make a large appearance in several slices.

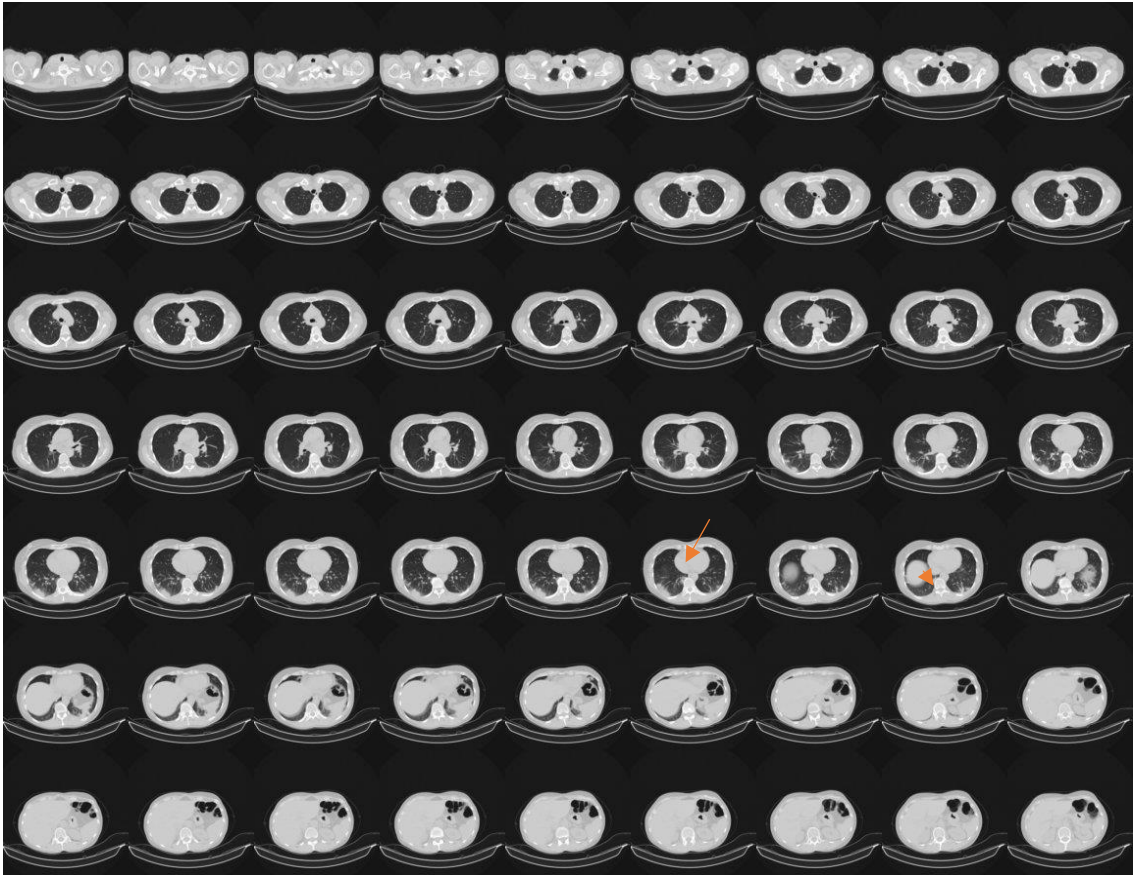

Figure 2. An axial view of a data volume in the form of montage. Arrow: heart. Arrow head: liver. This subject has confirmed diagnosis of COVID-19.

Hence, before the training, all images undertake pre-processing stage to remove the boundary, which is illustrated in Figure 3. Not only are the boundary of the lung removed but also the presence of the heart and liver, which leads to more focused content of interest. After this pre-processing stage by creating CT masks and segmentation, the final images have a resolution of 440 × 360. This task was completed using Matlab.

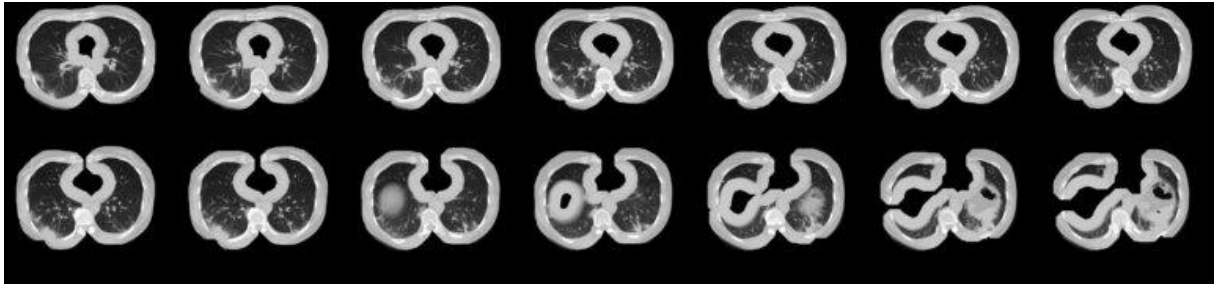

Figure 3. The images after pre-processing from the data volume in Figure 1, which are then applied to training or evaluation.

### 3.2 The challenges detecting COVID from Non-COVID CT images

Since there are still many unknowns regarding to COVID-19 virus, many biomarkers attributed to COVID-19 are not specific. The common visible patterns of COVID-19 include bilateral involvement and peripheral distribution, with superimposed interlobular septal thickening and visible intralobular lines. However, other patterns , such as with unilobar, perihilar patchy ground glass distribution do exist with COVID-19 patients [19].

Figure 4 demonstrates a group of images for a nonCovid patient, which present abnormal cloudy features.

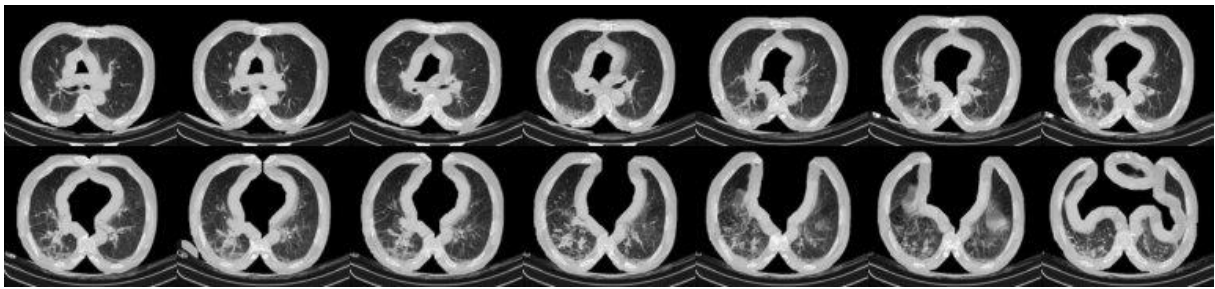

Figure 3. An example of nonCOVID dataset with cloudy patches.

### 4. Experimental results

The classification results are subject based, which is calculated from the predicted scores of all the 2D images for that subject. Considering the artefact that might be introduced during the pre-processing stage and a COVID volume contains nonCOVID 2D slices, the subject is classified as having COVID if more than a threshold (e.g. 25% ) number of slices are predicted as COVID. Similarly, if the remaining number, e.g. 75% or more, slices are predicted as nonCOVID, this subject will be classified as nonCOVID patient. Table 1 presents the confusion matrixes for the two deep learning systems, one is

COVID-CT system based on DenseNet [20] and one is Vision Transformer architecture shown in Figure 1. The evaluation results are based on validation dataset whlist the training employs the training datasets. Since the test data have no labels, they are not included in Table 2.

Table 2. Evaluation results for the two deep leaning system, CNN-DenseNet and VIT.

|  | CNN – DenseNet | | | | Vit—Vision transformer | | | |
| --- | --- | --- | --- | --- | --- | --- | --- | --- |
|  | COVID | Non COVID | Acc(%) | F1 | COVID | Non COVID | Acc(%) | F1 |
| COVID (predict) | 119 | 29 | 80.4 | 0.71 | 117 | 31 | 79.1 | 0.74 |
| Non COVID (predict) | 64 | 130 | 67.0 | 0.73 | 50 | 144 | 74.2 | 0.78 |
| **Average** | | | **73.7** | **0.72** | | | **76.6** | **0.76** |

Four predictions are submitted to this MIA-COV2019 competition with 2 thresholds for ViT (25% and 20%) and 2 for DenseNet (6% and 5%).

## 5. Conclusion

While participating MIA-COV19 competition, another aim of this work is to build an explainable system for medical application. Vision transformer architectures are built upon attention models and ae scalable when compared with CNN based models. Specifically, in the medical domain, the number of data sets can never be as large as current benchmark databases, e.g. ImageNet, with over millions of images. Hence a system that can still achieve good performance while employing limited number of datasets will make significant impact in the medical applications.

In comparison with CNN based model DenseNet for COVID-CT, ViT model appears to perform better with 76.6% accuracy whereas DenseNet realised accuracy of 73.7%.

While chest CT images are in 3D volume, it is a natural approach to process these data in 3D form. However, due to the large variations of slice numbers (depth), ranging from 4 to 1000+, with varying resolutions, generating 3D volumes present a challenge. Hence a depth of 32 slices, i.e. a volume of 224×224×32, is created for those subjects with sufficient depth images, by selecting slices evenly cross the whole volume. For example, if a 3D dataset has 64 slices in depth, then two sub-volumes are created for this subject with sub-volume 1 containing slices 1,3, .. 63 and sub-volume 2 having slices of 2,4, …, 64.

As address previously, the lesioned regions are proportionally small comparing to the whole volume, which might constitute the main reason that 3D based system perform far worse than 2D based models. Future work will further investigate this challenging issue. Another challenge remains to be the data volume size when preforming pre-pressing. Overall the training, validation and testing sizes are around 100 GB. Therefore pre-processing to segment lung content takes about 12 hours for all the subjects' dataset.